\documentstyle[prl,aps,epsfig,floats]{revtex}
\newcommand\beq{\begin{equation}}
\newcommand\eeq{\end{equation}}
\newcommand\bea{\begin{eqnarray}}
\newcommand\eea{\end{eqnarray}}

\begin{document}
\draft

\textheight=23.8cm
\twocolumn[\hsize\textwidth\columnwidth\hsize\csname@twocolumnfalse\endcsname

\title{An alternate model for magnetization plateaus in the molecular magnet 
$\rm V_{15}$}

\author{Indranil Rudra$^1$, S. Ramasesha$^1$ and Diptiman Sen$^2$}
\address{\it $^1$ Solid State and Structural Chemistry Unit,
Indian Institute of Science, Bangalore 560012, India \\
$^2$ Centre for Theoretical Studies,
Indian Institute of Science, Bangalore 560012, India}

\date{\today}
\maketitle

\begin{abstract}
Starting from an antiferromagnetic Heisenberg Hamiltonian for the fifteen
spin-$1/2$ ions in $\rm V_{15}$, we construct an effective spin Hamiltonian 
involving eight low-lying states (spin-$1/2$ and spin-$3/2$) coupled to a 
phonon bath. We numerically solve the time-dependent Schr$\ddot o$dinger 
equation of this system, and obtain the magnetization as a function of 
temperature in a time-dependent magnetic field. The magnetization exhibits 
unusual patterns of hysteresis and plateaus as the field sweep rate and 
temperature are varied. The observed plateaus are not due to quantum tunneling 
but are a result of thermal averaging. Our results are in good agreement with 
recent experimental observations.
\end{abstract}
\pacs{PACS numbers: ~75.45.+j, ~75.60.Ej, ~75.50.Xx}
\vskip.5pc
]

The synthesis of high-nuclearity transition metal clusters such as
${\rm Mn_{12},~Fe_8~and~V_{15}}$ \cite{synth} has provided an impetus to the
study of magnetism on the nanoscale. These transition metal clusters
are basically isolated transition metal complexes involving multi-dentate 
ligands; the chemical pathway between the metal ions in the transition metal 
complex dictates the nature of exchange interactions. The complex interplay of 
the topology of exchange interactions, magnetic dipolar interactions and
spin-lattice coupling has yielded a rich physics on the nanoscale which 
includes quantum tunneling, quantum phase interference and quantum coherence 
\cite{qtm,v15}. Quantum resonance tunneling is characterized by the 
observation of discrete steps or plateaus in the magnetic hysteresis loops 
at low temperatures. The signature of quantum interference is seen in the 
variation of the tunnel splitting as a function of the azimuthal angle of 
the transverse field for tunneling between $M_s~ =~ -10$ and $10-n$ states in 
molecular magnets with ground state spin-$10$ \cite{wernsdorferfe8}.
Quantum coherence-decoherence studies are important from the stand point
of application of these systems in quantum computations \cite{qcompu}. 

There have been several models proposed to understand these phenomena 
\cite{model}. Quantum hysteresis and interference have largely been studied 
by using an effective spin Hamiltonian with dipolar interactions with a time 
varying external magnetic field. The time evolution of the states of the 
system are carried out within a master-equation approach \cite{mastereq}. The 
decoherence phenomena has been studied by using a simple two state model in a 
transverse magnetic field \cite{decoherence}. Even though most of these 
clusters contain a fairly small number of metal ions, the spin on the metal 
ion, at least in the case of $\rm Fe_8$ and $\rm Mn_{12}$, is fairly large; a 
full quantum mechanical study of these systems is difficult because of the 
large Fock space dimensionalities of $1.69$ million and $100$ million 
respectively. However, the $\rm V_{15}$ cluster is far more amenable 
to a rigorous quantum mechanical analysis because of the much smaller
Fock space ($\approx 33,000$ dimensional) spanned by the unpaired spins of 
the system. A quantitative study of these systems requires at least the 
low-lying states of the full spin-Hamiltonian to be evolved in time quantum 
mechanically as the external magnetic field is ramped with time (as is done in 
experiments). In this letter, we study the magnetization of $\rm V_{15}$ by 
following its evolution as a function of a time-dependent magnetic field at
different temperatures. The low-lying states are obtained by solving the
exchange Hamiltonian corresponding to all the spins of the system.
A spin-phonon interaction is then introduced in the Hamiltonian. We thermally 
average the magnetization over the low-lying states after each of these states 
is independently evolved. We find that this model reproduces quantitatively 
all the experimental features found in the magnetization studies of $\rm 
V_{15}$ \cite{wernsdorferv15}.

The schematic structure of $\rm V_{15}$ is shown in Fig. 1. Structural and 
related studies on the cluster indicate that within each hexagon,
there are three alternating exchanges $J \approx 800~K$ which are 
the strongest in the system, and they define the energy scale of the problem. 
Besides, there are weaker exchange interactions between the spins involved 
in the strong exchange and also with the triangle spins which lie between the 
hexagons. All the exchange interactions are antiferromagnetic in nature.
The exchange pathways and their strengths \cite{wernsdorferv15} are also 
shown in Fig. 1. What is significant in the cluster is the fact that the 
spins in the triangle do not experience direct exchange interactions of any 
significance. The exchange Hamiltonian of the cluster
is solved using a valence bond basis in each of the total spin subspaces,
for all the eigenvalues. It is found that two spin-$1/2$ states and a 
spin-$3/2$ state are split-off from the rest of the spectrum by a gap of 
$0.6 J$ \cite{synth}. These eight states almost exclusively correspond 
to the triangle spins and they are the only states which will make significant 
contributions to sub-kelvin properties. We therefore set up an effective 
Hamiltonian in the Fock space of the three spins. We find that the form 
\bea
H_{sp-sp}~ =~ && \epsilon ~I ~+ ~\alpha ~(S_1 \cdot S_2 + S_2 \cdot S_3 + S_3 
\cdot S_1) \nonumber \\
&& +~ \delta ~S_1 \cdot S_2 \times S_3 ~,
\eea
where $\epsilon = -4.781808$, $\alpha =-0.001491$, and $\delta = 0.015144$ in 
units of the exchange $J$ (see Fig. 1), reproduces the low-lying eigenstates 
to numerical accuracy. Note the unusually large value of the three-spin 
interaction; this term has not been considered in the earlier literature
on this subject (such as Ref. [9]), and it is essential for the spin-$3/2$ 
state to lie between the two spin-$1/2$ states as is found to be the case in
this system.

The direct spin-spin interaction terms permitted by the $C_3$ symmetry are 
given by
\beq
H_{dip} ~=~ \gamma ~[~(S_{+}^3 + S_{-}^3) ~+~ i ~(S_{+}^3 - S_{-}^3)~] ~.
\eeq
We have also introduced a coupling between the spin states of the cluster and 
the phonons. The spin-phonon interaction Hamiltonian which preserves the $C_3$ 
symmetry is phenomenologically given by \cite{spph}
\bea
H_{sp-ph} ~=~ q ~ (b + b^\dagger) ~[~ && (S_+^2 + S_-^2) ~+~ i 
(S_+^2 - S_-^2) \nonumber \\
&& +~ (S_z^2 - {\frac{1}{3}} S^2) ~] ~,
\label{hspph}
\eea
where {\it q} is the spin-phonon coupling constant, $b$ ($b^\dagger$) is the 
phonon annihilation (creation) operator, and $\hbar \omega$ is the phonon 
frequency. For simplicity, we have assumed a single phonon mode although the 
molecule has various possible vibrational modes. The form of the interaction
in Eq. (\ref{hspph}) means that the phonons couple only to states with 
spin-$3/2$. We have restricted the dimensionality of the Fock space of the 
phonons to $15$ considering the low temperatures of interest.

The evolution of the magnetization as a function of the magnetic field has 
been studied by using the total Hamiltonian $H_{total}$, given by
\bea
H_{total} ~=~ && H_{sp-sp} ~+~ H_{dip} ~+~ H_{sp-ph} ~+~ \hbar \omega
(b^\dagger b + 1/2) \nonumber \\
&& +~ h_z (t) S_z ~+~ h_x (t) S_x ~,
\eea
where we have assumed that besides an axial field $h_z (t)$, a small 
transverse field $h_x (t)$
could also be present to account for any mismatch between the crystalline 
$z$-axis and the molecular $z$-axis. The numerical method involves setting up 
the Hamiltonian matrix in the product basis of the spin and phonon states 
$|i,j>$, where $|i>$ corresponds to one of the eight spin configurations of 
the three spins, and $j$ varies from $0$ to $14$, corresponding to the fifteen 
phonon states retained in the problem. The values we have assigned to the 
different parameters are $\gamma=10^{-3}$, $q=10^{-4}$ and $\hbar \omega =
1.25 \times 10^{-4}$, all in units of the exchange $J$ (see Fig. 1). 

To study the magnetization phenomena, we start with the direct product 
eigenstates of $H_{sp-sp}$ and $\hbar \omega (b^\dagger b + 1/2)$, and 
independently evolve each of the $120$ states $\psi_{ij}$ by using the time 
evolution operator
\beq
\psi (t + \Delta t) ~=~ e^{-i H_{total} \Delta t/\hbar} ~\psi(t) ~.
\eeq
The evolution is carried out in small time steps by applying the evolution
operator to the state arrived at in the previous step. The magnetic field
is changed step-wise in units of $0.015 T$. At each value of the magnetic 
field, the system is allowed to evolve for $300$ time steps of size $\Delta 
t$, before the field is changed to the next value. At every time step, the 
average magnetization $<M(t)>$ is calculated as
\bea
< M(t) > ~=~ \sum_{i=1}^8 ~\sum_{j=0}^{14} && ~
\frac{e^{-\beta [w_i+ h_{z}(t)m_i]}}{Z_{spin}} ~\frac{e^{-\beta \hbar \omega
(j + 1/2)}}{Z_{ph}}~ \nonumber \\
&& ~~~ <\psi_{ij} (t)| \hat{S_z} |\psi_{ij} (t)> ~,
\eea
where $w_i$ and $m_i$ are the eigenvalues and magnetizations of eigenstates of
$H_{sp-sp}$, $\beta = 1/k_B T$, $Z_{ph}$ is the phonon partition function of 
$\hbar \omega (b^\dagger b +1/2)$, and $Z_{spin}$
is the partition function of $H_{sp-sp}$ in the presence of the axial 
magnetic field. In Fig. 2, we show the energy level ordering of the effective 
spin Hamiltonian and the effect of the magnetic field on the eigenvalue 
spectrum. We also show the couplings between various states brought about by 
the magnetic dipolar terms and the spin-phonon terms; note that the spin-$1/2$ 
and spin-$3/2$ states are not connected to each other by these terms. 

In Fig. 3, we show the hysteresis plots of the system for different
temperatures. We see that at low temperatures, the plateaus in the
hysteresis plots are very pronounced. The plateau width at $<S_z> = - 0.5$ 
corresponds to $2.64$ T which is in excellent agreement with the experimental 
value of $2.8$ T \cite{wernsdorferv15} assuming that $J=800~K$.
We also find that the plateau vanishes above a temperature of $0.8$ K 
which is also in excellent agreement with the experimental value of $0.9$ K 
\cite{wernsdorferv15}. The inset in figure shows the temperature variation 
of the plateau width. We note that the plateau width falls off rapidly with 
temperature, and an exponential fit to $W~=~A~\exp (-T/\Omega)$ 
(see Fig. 3) gives the characteristic temperature $\Omega$ 
to be $0.2$ K. This small value of $\Omega$ is because there are no large
barriers between the different magnetization states in this system, unlike
the high spin molecular magnets such as $\rm Mn_{12}$ \cite{model}.

We also observe that when the field is swept more rapidly, there are 
additional plateaus at intermediate values of magnetization. For example, in 
Fig. 4 the field sweep rate is increased by a factor of five compared to 
Fig. 3, and we find a small plateau of width $0.03 ~T$ near $H=0.15 ~T$
at a value of $< S_z > = -0.375$. This is because near that field, some of
the spin-$3/2$ states become degenerate in energy; subsequently, as the 
magnetic field is increased, the system stays locked 
in some of those states if the sweep rate is too high. This plateau vanishes 
on warming the system slightly. We have also studied the effect of cycling
the field on the plateaus. In Fig. 5, we show the magnetization as a
function of field at different temperatures and sweep rates. It is 
interesting to see that the system does not have either remnance or 
coercivity; all the hysteresis plots pass through the origin. The effect of 
varying the rate of scanning the field is also shown in Fig. 5. 
We find that as the scanning rate increases, the hysteresis in the plot of
magnetization {\it vs} field decreases, and the plateau feature is
almost identical in both the scanning directions. This could be 
because of the slow relaxation of the magnetization which is indeed the
reason why the plateaus occur in the first place. We also find that the 
transverse field term does not affect any of our results significantly.

To summarize, we have derived an effective Hamiltonian from the
exchange Hamiltonian of the full $\rm V_{15}$ system. In the presence of a 
time varying magnetic field, the states of the effective Hamiltonian are 
allowed to evolve under the influence of magnetic dipolar interactions and 
a spin-phonon coupling. During the time evolution, the magnetization is 
followed as a function of the applied magnetic field. The calculated
$M$ {\it vs} $H$ plots show magnetization plateaus at low temperatures. The 
width of the plateau at low temperature as well as the temperature at which
the plateau vanishes are in excellent agreement with experimental values. 
It is also shown that the number of plateaus observed depends upon the
scanning speed of the magnetic field. When the magnetic field is cycled, the 
hysteresis plots pass through the origin indicating the absence of remnance 
and coercion. The hysteresis is pronounced for slow scanning speeds. From our 
results, it appears that the magnetization plateaus in $\rm V_{15}$ is not a 
consequence of quantum resonant tunneling but is a result of thermal 
averaging. We also find that the magnetization does not show any oscillation 
with time during evolution indicating the absence of quantum tunneling.

We thank the Council of Scientific and Industrial Research, India for their 
grant No. 01(1595)/99/EMR-II.

\begin{figure}
\begin{center}
\epsfig{figure=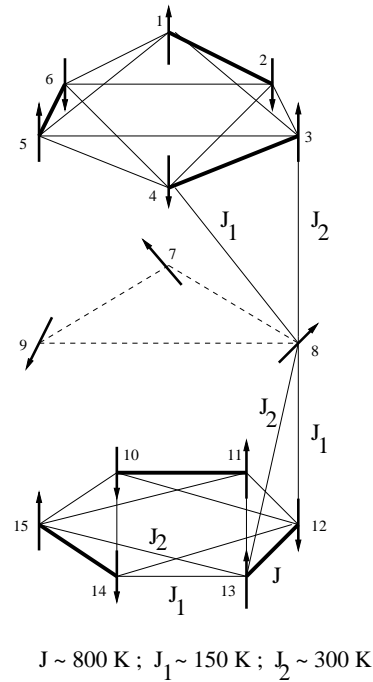,height=9cm}
\end{center}
\caption{Schematic exchange interactions in a $\rm V_{15}$ cluster. There is 
no direct exchange interaction amongst the triangle spins. Interactions not 
shown explicitly can be generated from the $C_3$ symmetry of the system.}
\end{figure}

\begin{figure}
\begin{center}
\epsfig{figure=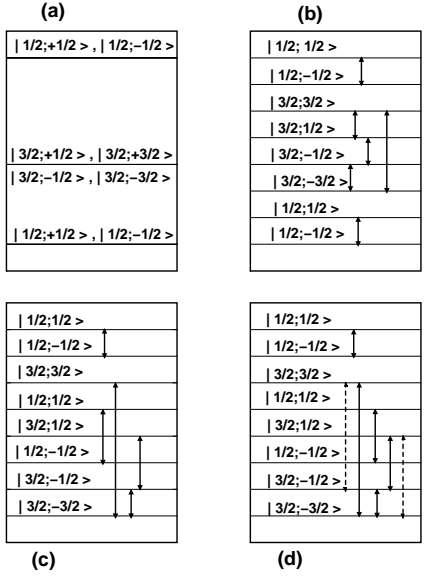,bbllx=50,bblly=0,bburx=500,bbury=800,height=9cm}
\end{center}
\caption{(a) Eigenstates of the effective spin Hamiltonian
$H_{sp-sp}$, (b) Eigenstates in the presence of a moderate axial field.
Arrows show the states connected by the dipolar terms and the transverse
field. (c) is the same as (b) but in a stronger field, (d) describes the
effect of spin-phonon terms (shown by arrows with broken lines) on (c).}
\end{figure}

\begin{figure}
\begin{center}
\epsfig{figure=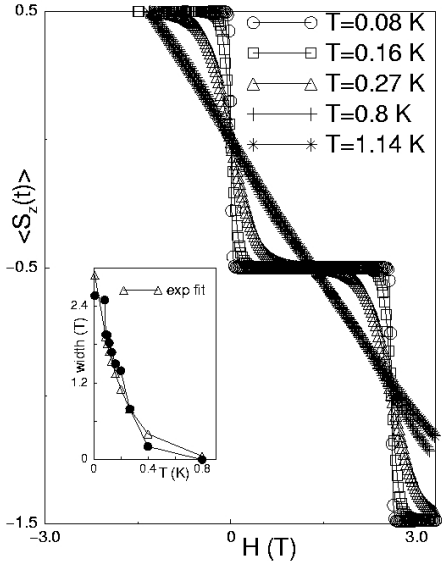,bbllx=50,bblly=0,bburx=500,bbury=800,height=9cm}
\end{center}
\caption{Plot of magnetization {\it vs} axial field at different temperatures. 
Inset shows plateau width as a function of temperature (full circles). 
Triangles in the inset correspond to the values from the fit to an exponential 
function.}
\end{figure}

\begin{figure}
\begin{center}
\epsfig{figure=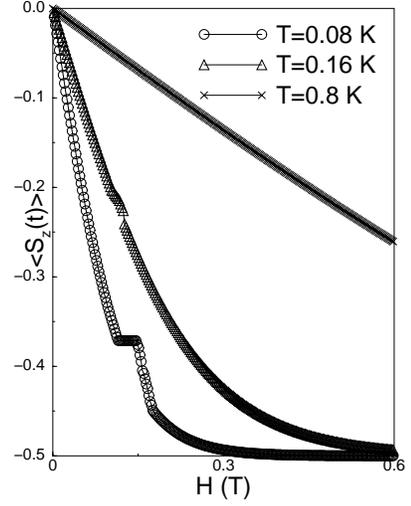,bbllx=50,bblly=0,bburx=500,bbury=800,height=9cm}
\end{center}
\caption{Magnetization as a function of axial field for a faster sweep rate at 
three different temperatures.}
\end{figure}

\begin{figure}
\begin{center}
\epsfig{figure=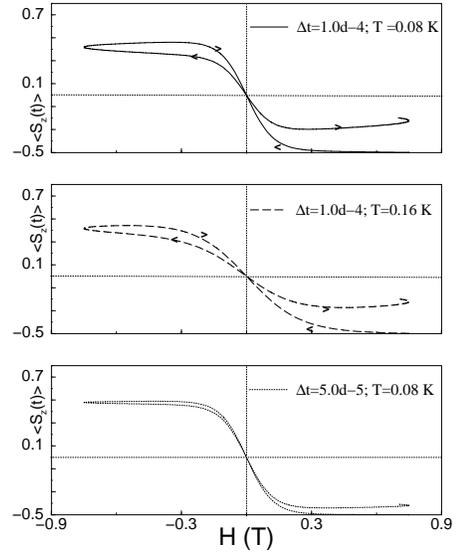,bbllx=50,bblly=0,bburx=500,bbury=800,height=9cm}
\end{center}
\caption{Magnetization {\it vs} axial field for a full cycling of the field at 
different temperatures and sweep rates.}

\end{figure}

\end{document}